\documentclass[twocolumn]{aastex6}

\usepackage[utf8]{inputenc}
\usepackage{natbib}
\usepackage{graphicx}
\usepackage{subfigure}
\usepackage{soul}
\graphicspath{{./figs/}}

\usepackage{multirow, amsmath}

\begin{document}

\title{The unrelaxed dynamical structure of the galaxy cluster Abell 85 }
\author{Heng Yu\altaffilmark{1,2,3}, Antonaldo Diaferio\altaffilmark{2,3}, Irene Agulli\altaffilmark{4,5,2,3}, 
J. Alfonso L. Aguerri\altaffilmark{4,5}, Paolo Tozzi\altaffilmark{6,1}}

\altaffiltext{1}{Department of Astronomy, Beijing Normal University,
    Beijing, 100875, China}
\altaffiltext{2}{Dipartimento di Fisica, Universit\`a di Torino, Via
  P. Giuria 1, I-10125 Torino, Italy}
\altaffiltext{3}{Istituto Nazionale di Fisica Nucleare (INFN), Sezione di Torino, Via
  P. Giuria 1, I-10125 Torino, Italy}
\altaffiltext{4}{Instituto de Astrofisica de Canarias, C/Via Lactea s/n, E-38200 La Laguna, Tenerife, Spain}
\altaffiltext{5}{Departamento de Astrofisica, Universidad de La Laguna, E-38205 La Laguna, Tenerife, Spain}
\altaffiltext{6}{INAF—Osservatorio Astrofisico di Arcetri, Largo E. Fermi, I-50122 Firenze, Italy}


\begin{abstract}

For the first time, we explore the dynamics of the central region of a galaxy cluster
within $r_{500}\sim 600h^{-1}$~kpc from its center by combining optical and X-ray spectroscopy. 
We use (1) the caustic technique that
identifies the cluster substructures and their galaxy members with optical spectroscopic data, 
and (2) the X-ray redshift fitting procedure that estimates the redshift distribution
of the intracluster medium (ICM). 
We use the spatial and redshift distributions of the galaxies and of the X-ray emitting gas to
associate the optical substructures to the X-ray regions.
When we apply this approach to Abell 85 (A85), 
a complex dynamical structure of A85 emerges from our analysis: 
a galaxy group, with redshift $z=0.0509 \pm 0.0021$ 
is passing through the cluster center along the line of sight dragging part of the ICM present in the 
cluster core; two additional groups, at redshift $z=0.0547 \pm 0.0022$ and $z=0.0570 \pm 0.0020$, are going through the cluster
in opposite directions, almost perpendicularly to the line of sight, and have substantially perturbed the dynamics of the ICM. An additional group in the
outskirts of A85, at redshift $z=0.0561 \pm 0.0023$, is associated to a secondary peak of the X-ray emission,  at redshift $z=0.0583^{+0.0039}_{-0.0047}$.
Although our analysis and results on A85 need to be confirmed by high-resolution spectroscopy, they 
demonstrate how our new approach can be a powerful tool to constrain the formation history of galaxy clusters by unveiling their 
central and surrounding structures.

\end{abstract}

 
\section{Introduction}

Optical and X-ray observations of clusters of galaxies and their environs support the scenario where 
clusters form by the accretion of matter from their surroundings \citep{2001Rines,2013Medezinski,2015Eckert}, 
as implied by the distribution of galaxies in large and dense galaxy redshift surveys 
\citep{1986Lapparent,1989Geller,2001Colless,2003SDSS,2009SDSS7,2014Ahn,2016Hwang}
and expected in hierarchical clustering models \citep{1996Bond,2005Colberg}.

There are various dynamical signatures of this mass accretion process: (i) the presence of substructures in the galaxy 
density distribution of the cluster
on the sky and in redshift space \citep[e.g.,][and references therein]{1982Geller,1988Dressler,2007Ramella,2015Grillo,2015Girardi,2016Balestra}; 
(ii) the clumpy distribution either of the hot intracluster medium (ICM) observed in the X-ray band \citep[e.g.,][]{1993Mohr,2001Kolokotronis,2005Jeltema,2010Zhang,2015Parekh}
or (iii) of the dark matter distribution inferred from gravitational lensing effects 
\citep[e.g.,][]{1996Kneib,2000Hoekstra,2014Okabe}; 
(iv) 
the presence of diffuse radio emission with elongated or peculiar morphologies 
\citep[e.g.,][and references therein]{2016Girardi}. 

The dynamically unrelaxed state of the cluster might 
also substantially affect the location of the brightest cluster galaxy \citep[BCG;][]{2014Lauer}, that we 
expect at the bottom of the gravitational potential well of the cluster \citep{2004Lin}: 
the BCG can be displaced from the peak of the projected galaxy density distribution \citep{1983Beers}, from the peak of the X-ray emission 
\citep[e.g.,][and references therein]{2016Rossetti}, and from the global redshift of the cluster \citep{1991Beers,2012Zitrin}.

Identifying cluster substructures and assessing their properties is thus a crucial tool to probe the mass
assembly of cosmic structures. Moreover, it can substantially contribute to the investigation of the effect of 
environment on the evolution of galaxy properties 
\citep[e.g.,][]{2012Hwang, 2013Pranger,2014Pranger,2015Hess, 2015Lee, 2016aAgulli, 2016Utsumi}. 

Substructures are relevant for an additional reason. On the scale of galaxies, 
significant discrepancies between observations and the cold dark matter model 
emerge, namely 
the missing satellite problem, the too-big-to-fail problem, the angular momentum catastrophe and the cusp-core problem 
\citep[see, e.g.,][for a review]{2016Popolo}. 
A possible solution to these discrepancies is the adoption of additional dark matter components, like
warm, self-interacting, or interacting dark matter particles 
\citep[e.g.,][]{2000Spergel,2013Viel,2014Boehm}.  
Although still debated, investigating cluster substructures could in principle distinguish among these dark matter variants. In fact,
ordinary matter and different kinds of dark matter behave differently during the collision between
cluster components: this different behavior might produce observable differences 
\citep[e.g.,][]{2014MNRAS.437.2865K,2015Sci...347.1462H,2015Massey,2016Robertson}. 

The investigation of the substructure properties requires objective methods to identify cluster substructures. These
methods are based on optical or X-ray data, as very briefly reviewed in  \citet{Yu2015}. 
Investigations that combine more than one approach are numerous
\citep[e.g.,][]{1996AJ....112.1816M,2011Bourdin,2014Guennou,2014ApJ...797..106H,2015Jauzac,2015ApJ...812..153O,2016Jauzac,2016ApJ...819..113O}, and are, in fact, highly desirable to
infer, more robustly, the assembly history of the cluster. 
In addition, and equally important, the combination of different methods can assess 
the systematic errors of the methods themselves \citep{2013Geller,2014Geller}.

Here, for the first time, we combine optical and X-ray spectroscopy to associate substructures
in the galaxy density distribution to the clumps of the ICM. 

With data in the optical band, substructures are difficult to identify even when spectroscopic information is available; 
in fact, substructures usually contain a limited number of galaxies,
and it is extremely difficult to assess the membership of their galaxies, mostly because of 
the confusion introduced by projection effects. 
Here we apply the caustic technique, that is
based on spectroscopic redshifts \citep{1997Diaferio, 1999Diaferio,Serra2011},
to identify the members of the cluster core and
of the cluster substructures and estimate the redshifts of these structures. \citet{Yu2015} 
apply this technique to a sample of 150 mock catalogs of clusters extracted from N-body simulations and show 
that this technique can identify catalogs of substructures that are at 
least $\sim 60$\% complete and contain at most $\sim 50$\% spurious substructures. No other available technique 
appears to perform better on realistic mock cluster catalogs.

When the X-ray emission of clusters is bright enough, the X-ray spectrum may also be used 
to estimate the redshift of the emitting gas \citep{2011Yu}, although 
this estimate may suffer from the limited energy resolution of the X-ray detector.
\citet{2015Liu} and \citet{2016Liu} outlined a simple and effective technique to measure the
projected X-ray redshifts in different ICM regions and assess the
statistical and systematic errors on these redshifts.

By combing the caustic technique and the X-ray redshift fitting, we can infer the 
motion of the cluster substructures and unveil the complex dynamics of the cluster. 
We apply this method to a specific cluster: Abell 85 (A85). This cluster
is the only system currently available where the sample of galaxy spectra is large enough that we can
identify substructures in the same central region
covered by the X-ray spectroscopy. We show how, with our approach, we can infer the recent accretion
history of A85. It thus appears to be feasible that, when applied to a large sample of clusters with high-quality spectra,
our analysis can directly probe the mass assembly of clusters, provide further constraints on hierarchical
clustering scenarios on small scales, and eventually probe the properties of the dark matter particles. 

In section \ref{sec:A85}, we review the
estimates of the redshift of A85 with optical and X-ray spectroscopy.  Sections \ref{sec:sdata} and \ref{sec:odata} present 
the optical data we use here and their analysis; the calibration 
and spectrum fitting of the X-ray data is illustrated in Section \ref{sec:xdata}. 
The analysis combining optical and X-ray spectroscopy is described in Section \ref{sec:comb}.
Our conclusions and prospects are given in Section \ref{sec:resu}.

\section{The global redshift of A85} 
\label{sec:A85}

The redshift of a galaxy cluster is usually estimated from the distribution
of the redshifts of its member galaxies.
The redshift of the brightest galaxy is also adopted 
when the spectroscopic data are insufficient. 
However, unavoidably, the estimate of the cluster redshift is affected by the radial velocity of the substructures of
the cluster and therefore by its dynamical state.
Therefore, the measurement of the cluster redshift is far from being a trivial issue and is 
deeply connected with the study of the cluster dynamics and the presence of substructures.

The nearby cluster A85 is a perfect case for a test of our approach combining optical and X-ray spectroscopy. 
A85 is a rich cluster with a BCG (MCG-02-02-086), 
X-ray substructures \citep{2010Tanaka,2014Schenck,2015Ichinohe},
and filaments \citep{2003Durret,2008Boue}.
The measured redshift $z$ of A85 is slightly different according to different analyses.
In early studies, \citet{1989Abell} measure a value of $z=0.0518$ with 59 member galaxies.
\citet{1991Struble}, in their compilation
of redshifts for Abell clusters, use 116 galaxies to derive the redshift $z=0.0556$.
\citet{1998Durret} perform much deeper observations 
and measure $z=0.0555 \pm 0.0003$ with 305 optical member galaxies, 
whose velocities are in the range 13,350-20,000 km~s$^{-1}$. 
\citet{2001Oegerle} estimate a redshift of $z=0.0551 \pm 0.0003 $ with 130 member galaxies.
The NOAO Fundamental Plane Survey reports $z=0.0547 \pm 0.0004$ with 58 member galaxies \citep{Smith2004}.
\citet{2005Miller} confirm the redshift of \citet{1991Struble} $z=0.0556$ with 82 member galaxies 
from the SDSS survey Second Data Release (DR2), similar to $z=0.0557$ found  
with 191 members by \citet{Rines2006} in the CIRS survey.
\citet{2007Aguerri} use 273 redshifts of SDSS-DR4 to derive $z=0.0555 \pm 0.0001$.
\citet{2009Bravo} find 367 member galaxies with a compilation of 
Abell cluster member galaxies \citep{2005Andernach}, and derive $z=0.0553 \pm 0.0002$.

\begin{figure}[htbp]
\includegraphics[width=0.45\textwidth]{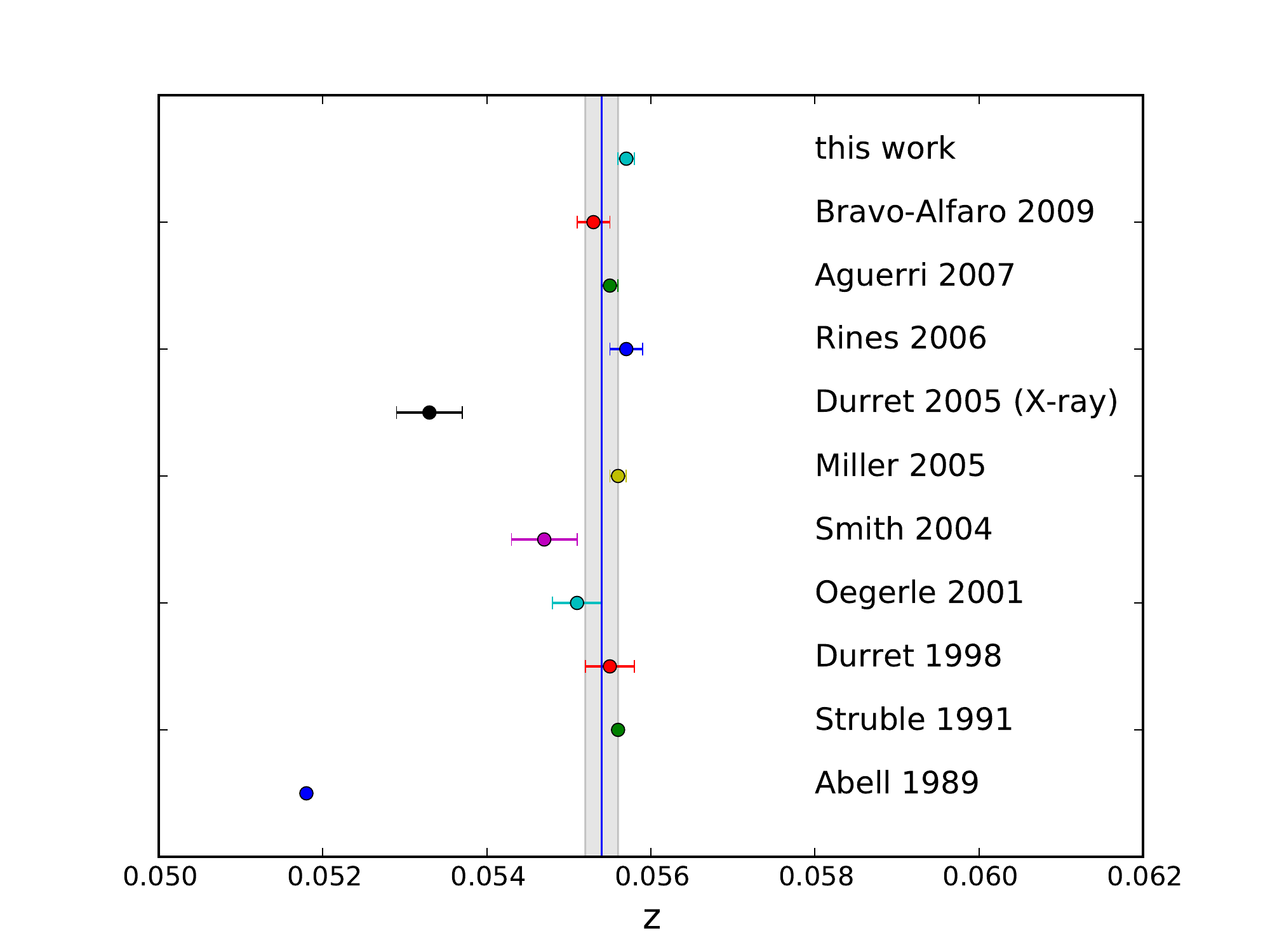}
\caption{The measured redshift of A85 in the literature. 
The blue vertical line and the grey shaded area indicate the redshift of the BCG and its error, 
$z=0.0554 \pm  0.0002$ \citep{2008DR6}.}
\label{fig:zlist}
\end{figure}

Figure \ref{fig:zlist} shows how some of the 
redshifts listed above significantly differ from the redshift of the
brightest galaxy $z=0.0554 \pm  0.0002$ \citep{2008DR6}; in addition, 
some differences among the redshifts 
are larger than their quoted errors:
the redshift tends to be underestimated when the sample is not large enough, and,
in fact, the disagreement has been recently alleviated by deep redshift surveys like SDSS.
Figure \ref{fig:zlist} shows that the most relevant discrepancy appears between the
redshift determined from the X-ray data \citep{2005Durret} and the optical redshifts, if we
neglect the early redshift estimate by  \citet{1989Abell}.
In fact, with the {\sl XMM-Newton} observations, 
\citet{2005Durret} find that the X-ray redshift of the core of A85, $z=0.0533 \pm 0.0004$, 
is significantly smaller than the average redshift $z=0.0557$.
\citet{2005Durret} suggest that this discrepancy is originated by the presence of cluster substructures.
This suggestion remains unproved. 

Here, we use the largest data sample and the newest analysis techniques to 
unveil the complex structures of A85 that 
explains the aforementioned discrepancy in the redshift measurements.

\section{Optical Spectroscopic Data}
\label{sec:sdata}

We use the spectroscopic redshift catalog compiled by \citet{Agulli2014,2016Agulli} 
based on data from SDSS-DR6 \citep{2008DR6}, the VIsible Multi-Object Spectrograph at the Very Large Telescope (VI-
MOS@VLT, Program 083.A-0962(B), PI R. S\'anchez-Janssen, 2009 August), 
the William Herschel Telescope (WHT), and the NASA/IPAC Extragalactic Database (NED).
\footnote{ The redshifts of the cluster members are publicly available in \citet{2016Agulli}.}

Our final sample contains 1603 galaxies: 
281 galaxies are brighter than $m_r = 17.77$, where $m_r$ is the de-red SDSS-DR6
red-band magnitude, and the sample is 95\% complete to this magnitude; the remaining 1322 galaxies have
fainter magnitudes down to $m_r = 22$.

We have 241 redshifts from SDSS-DR6 
with spectral resolution in the range $R=[1850,2200]$, yielding an uncertainty $\sim 50$~km~s$^{-1}$ on the Hubble radial velocity derived from the Doppler shift. 
The 1294 redshifts  from  VLT/VIMOS and 19 from WHT are derived with low-resolution spectra ($R=180$ and $R=280$ 
from VLT/VIMOS and WHT, respectively); these resolutions yield an uncertainty $\sim 500$~km~s$^{-1}$ on 
the radial velocity, as demonstrated by a number of repeated observations \citep{Agulli2014}. 
We complete our spectroscopic redshift sample with  49 redshifts from the NED database, for which
we do not know the radial velocity uncertainty. To be conservative, 
we assume $\sigma_{sp}=500$~km~s$^{-1}$ as the uncertainty on each individual redshift of our sample.

The large uncertainty on the redshifts and the incompleteness of the galaxy
sample at the faint end might affect our 
substructure identification, as we will discuss in section \ref{sec:resu}. 
However, our aim here is to provide an example of 
what information we can extract by 
comparing the redshift distribution
of galaxies with the redshift distribution of the X-ray emitting gas. 

Our galaxy sample is the only one currently available
that is dense enough in the central region of the cluster
that is covered by the X-ray spectroscopy, namely within $r_{500}$, or $\sim 0.6 h^{-1}$~Mpc for
A85 \citep{Rines2006}. The combination of the Chandra fields of A85 probes
a box $\sim (0.60\times 1.2) h^{-2}$~Mpc$^2$ around the X-ray peak; this area roughly corresponds to the field
centered on $[\alpha,\delta]=[10.464623,-9.3699074]$~deg with $0.1$~deg extension in right ascension
$\alpha$ and $0.2$~deg extension in declination $\delta$ (see Figure \ref{fig:sky} below). 
In this area, we have a unique sample of $ 243$ redshifts, 
out of which $171$ are cluster members, as we will show below.

In addition, the X-ray spectroscopy also yields uncertainties of at least $400$~km~s$^{-1}$, comparable
to the opical uncertainty. Therefore, both optical and X-ray redshifts are insensitive 
to substructures with redshift deviations smaller than $\sim 500$~km~s$^{-1}$.

\section{Substructures in the galaxy distribution}
\label{sec:odata}

\begin{figure}[htbp]
\includegraphics[width=0.45\textwidth]{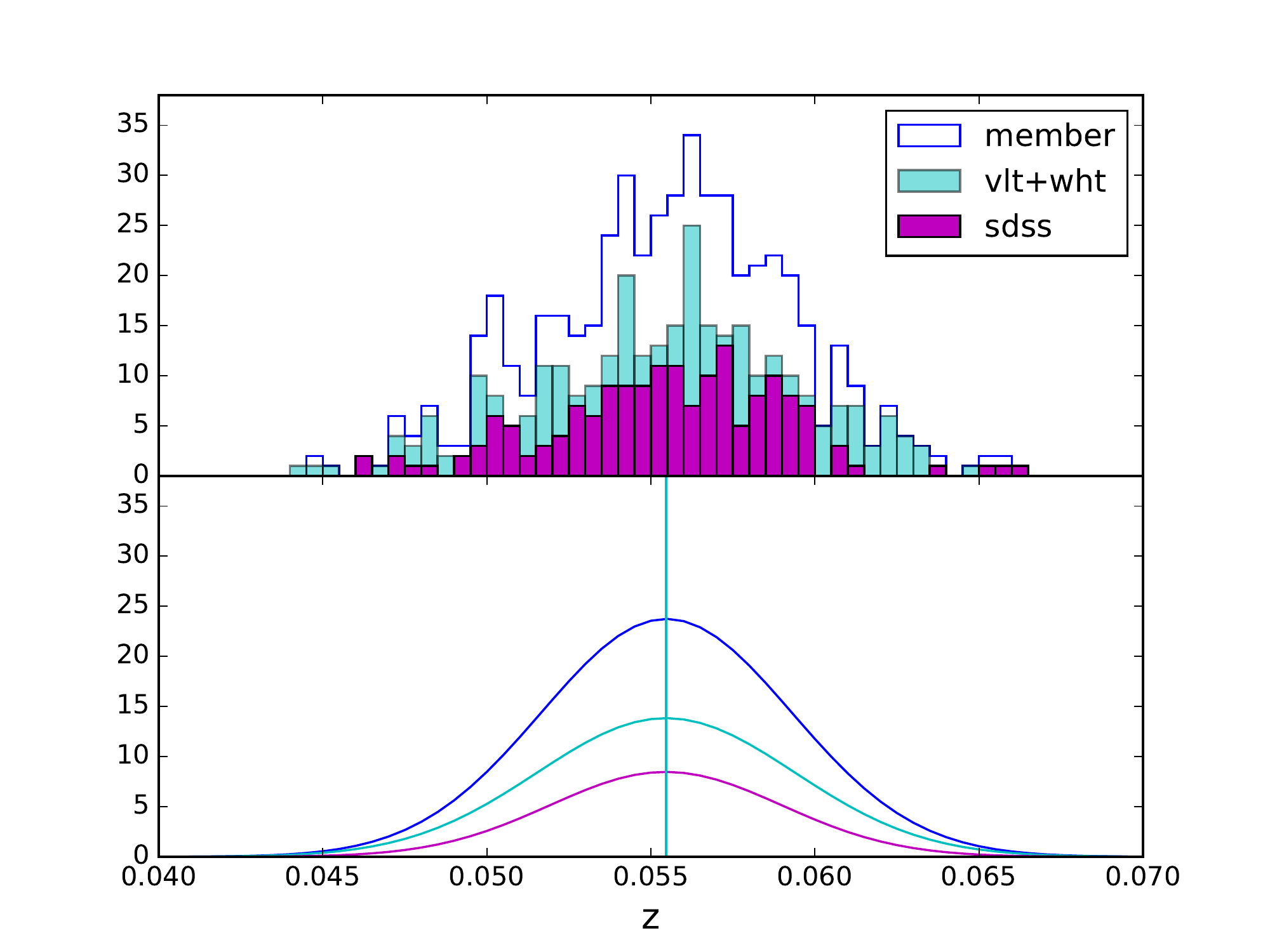}
\caption{The velocity histogram of the 511 member galaxies in the range $0.04-0.07$ . 
The blue open histogram shows the entire distribution.
The cyan bars show the 307 galaxies from VLT and WHT. 
The magenta bars show the 169 galaxies from SDSS.
The curves in the lower panel show the Gaussian fit after 3$\sigma$ clipping.
The vertical lines indicate the three Gaussian centroids that overlap each other.}
\label{fig:numhst}
\end{figure}

There are 515 redshifts in the range $z=[0.04,0.07]$. 
The distribution of these redshifts $z$ is shown by the open histogram in the upper panel of Figure \ref{fig:numhst}.
The application of the traditional $3\sigma$ clipping procedure only removes 4 galaxies and 
returns 511 cluster members. 
The mean redshift and the redshift dispersion of these members are 0.0555 and 0.0038, 
respectively. The lower panel of Figure \ref{fig:numhst} shows that 
our individual subsamples have comparable average redshifts and velocity dispersions: the VLT+WHT 
and SDSS galaxy samples have velocity dispersion $1179$~km~s$^{-1}$ and $1060$~km~s$^{-1}$, respectively. 

The histogram is skewed towards the left,
suggesting an ongoing merging process that might be one of the 
reasons for the discrepancy among some of the redshifts reported in the literature. 

The caustic method \citep{1997Diaferio, 1999Diaferio,Serra2011} is based on 
optical spectroscopic data and has been proved to be a reliable procedure
to identify cluster members \citep{2013Serra}.
It also is a promising tool to identify substructures \citep{Yu2015}.
It uses the relative binding energy to link galaxies in the field 
of view and arrange them in a binary tree. By tracing nodes that contain 
the largest number of galaxies at each bifurcation, we can draw 
the main branch of the tree. When walking along the main branch 
from the root to the leaves, the velocity dispersions  $\sigma$ of the leaves of each node 
settles onto a $\sigma$ plateau.  The two boundaries of the plateau identifies two thresholds
that are used to cut the tree at two levels: the first level identifies
the cluster members, the second level identifies the cluster substructures
\citep[see][for further details]{2013Serra,Yu2015}.

\begin{figure}[htbp]
\includegraphics[width=0.45\textwidth]{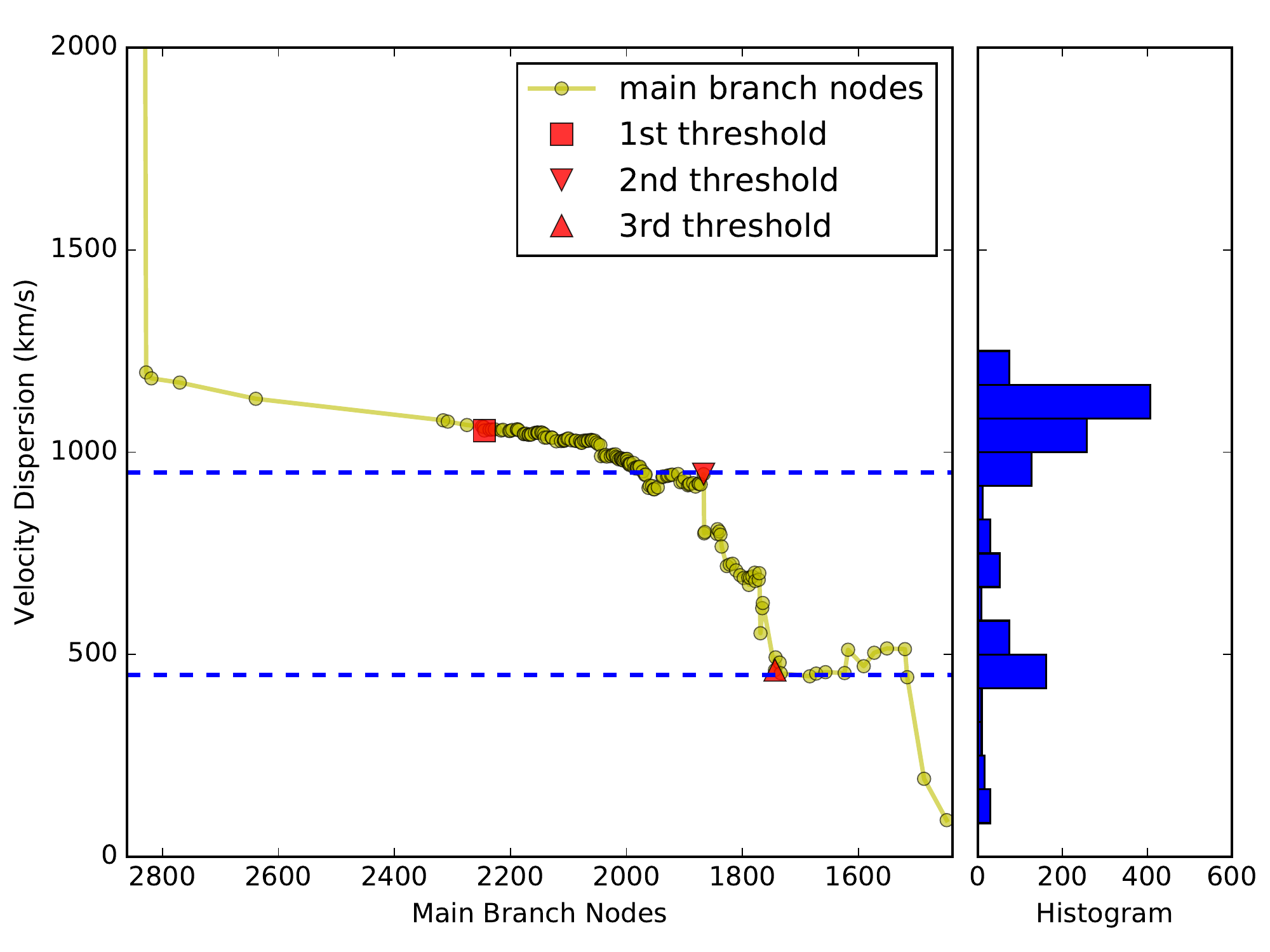}
\caption{The velocity dispersion of the leaves of each node along the
main branch of the binary tree of A85. The histogram in the right panel
shows the node numbers in the different velocity dispersion bins. 
The blue dashed lines indicate the $\sigma$ plateaus.
The red symbols are the selected thresholds.}
\label{fig:sp}
\end{figure}

Figure \ref{fig:sp} shows the velocity dispersion along the main branch of the binary tree of A85. 
The 494 galaxies hanging below the first threshold (red solid square in Figure \ref{fig:sp}) are the cluster members 
identified by the $\sigma$ plateau:{\footnote{The actual list
of cluster members provided by the caustic technique is determined by the location of the
caustics in the redshift diagram of the cluster \citep{2013Serra}. For the
sake of simplicity, we omit this second step in this analysis: 
the general conclusions of our analysis remain unaffected.}}
their average redshift is $z_{avg}=0.0554$ and their redshift dispersion $v_{\rm disp}$ is 0.0035,
as listed in the first row of Table \ref{table:subs}. This velocity dispersion, $1054$~km~s$^{-1}$, is larger
than the velocity dispersion $692^{+55}_{-45}$~km~s$^{-1}$ found by \citet{Rines2006} based on the 191 galaxy members brighter
than $m_r=17.77$ from the SDSS-DR4 catalog. These galaxies were identified 
from the location of the caustics
in the redshift diagram extending to $10 h^{-1}$~Mpc from the cluster center. 
Our sample is slightly different: it covers an area of $3.0\times 2.6 h^{-2}$~Mpc$^2$, 
and contains all the galaxies on the main branch hanging from the first threshold, 
thus including those galaxies 
that are not within the caustics in the redshift diagram and mostly in the
tails of the velocity distribution. Our sample limited to the SDSS galaxies within the caustics
yields a velocity dispersion consistent with \citet{Rines2006} result.
 
\begin{figure*}[htbp]
\centering
\includegraphics[width=0.85\textwidth]{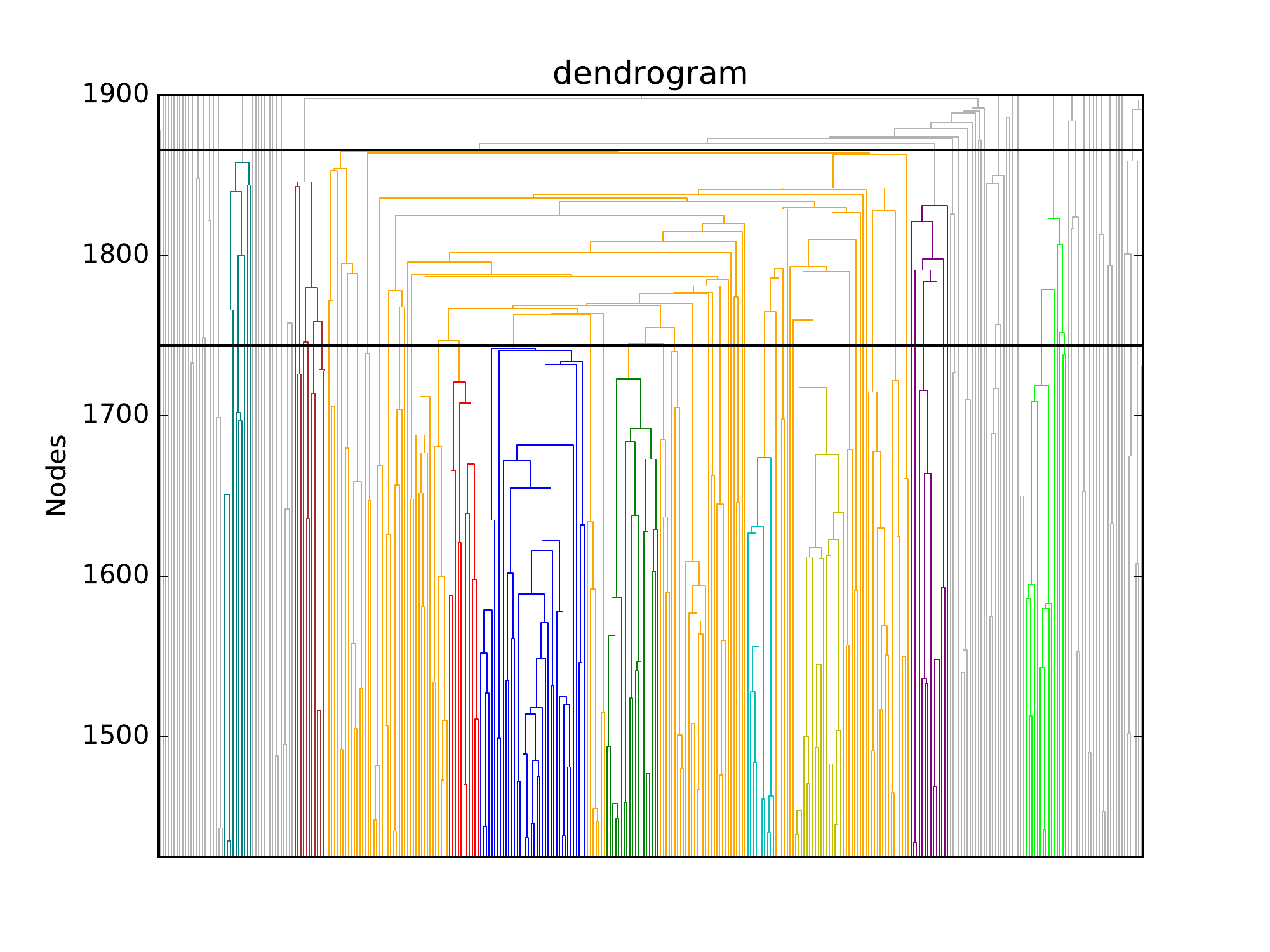}
\caption{The central part of the A85 dendrogram, including the 350 most bound galaxies. 
The two solid horizontal lines are the 2nd and 3rd thresholds that cut the tree 
at two different levels; the first threshold cuts the tree at a higher level  
of the dendrogram that is not shown here. The galaxies are at the bottom of the dendrogram 
and the vertical lines show how the binary tree links them hierarchically. 
The colors indicate different structures. 
The  orange structure corresponds to sub0 listed in Table \ref{table:subs} identified by the
second threshold. The 3rd threshold breaks it into 5 smaller substructures and individual galaxies; 
we set to 10 the minimum number of galaxies that defines a substructure.}
\label{fig:tree}
\end{figure*}

Figure \ref{fig:sp} shows two plateaus on the main branch: around 950, 
and 450 km~s$^{-1}$.
The presence of more than one plateau highlights the complex dynamical state of A85.
The first plateau around 950 km~s$^{-1}$ is the  $\sigma$ plateau automatically identified
by the caustic algorithm,
whose first boundary identifies the cluster members.
The second boundary of the $\sigma$ plateau sets the second threshold that identifies 
the cluster substructures.  
\citet{Yu2015} show that the $\sigma$ plateau automatically located by the algorithm does not 
always return the most appropriate threshold for the identification of the cluster substructures, 
especially when the cluster dynamical state is particularly complex. 
In these cases, as the substructure threshold, we can pick the starting 
node of a plateau, or alternatively, the first node below the previous plateau. 
In our case, we choose an additional threshold 
below 950 km~s$^{-1}$, the red triangle around 450 km~s$^{-1}$ shown in 
Figure \ref{fig:sp} (the third threshold hereafter). 
By using these thresholds, we can explore how galaxies populate the cluster at 
different hierarchical levels. Figure \ref{fig:tree} zooms into the central part of
the binary tree of our full sample: the upper and lower horizontal lines are the
2nd and 3rd thresholds and individual substructures are identified by different colors. 

\begin{figure*}[htbp]
\subfigure[threshold 2]{\includegraphics[trim = 60 0 60 0,clip,width=0.45\textwidth]{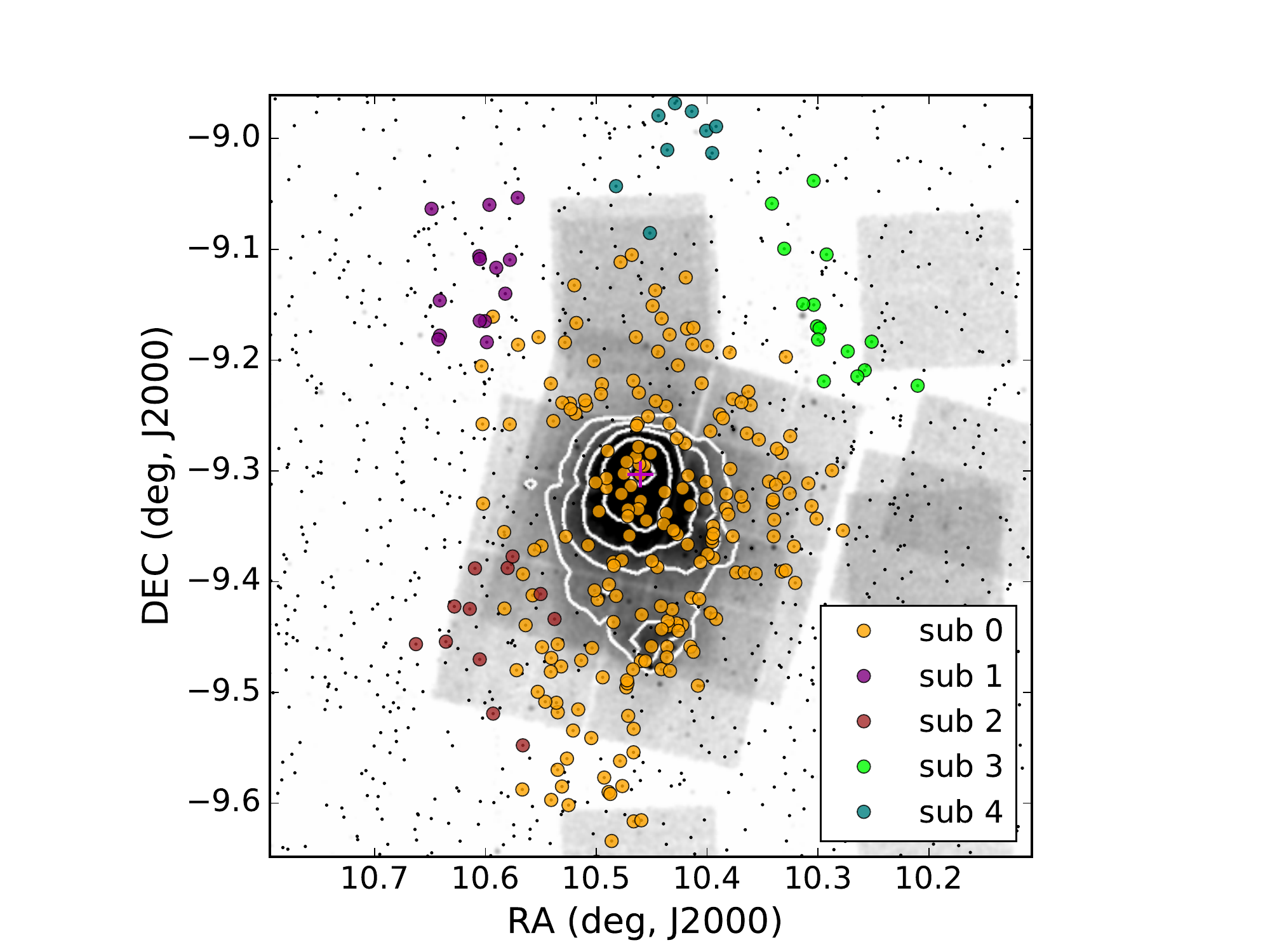}}
\subfigure[threshold 3]{\includegraphics[trim = 60 0 60 0,clip,width=0.45\textwidth]{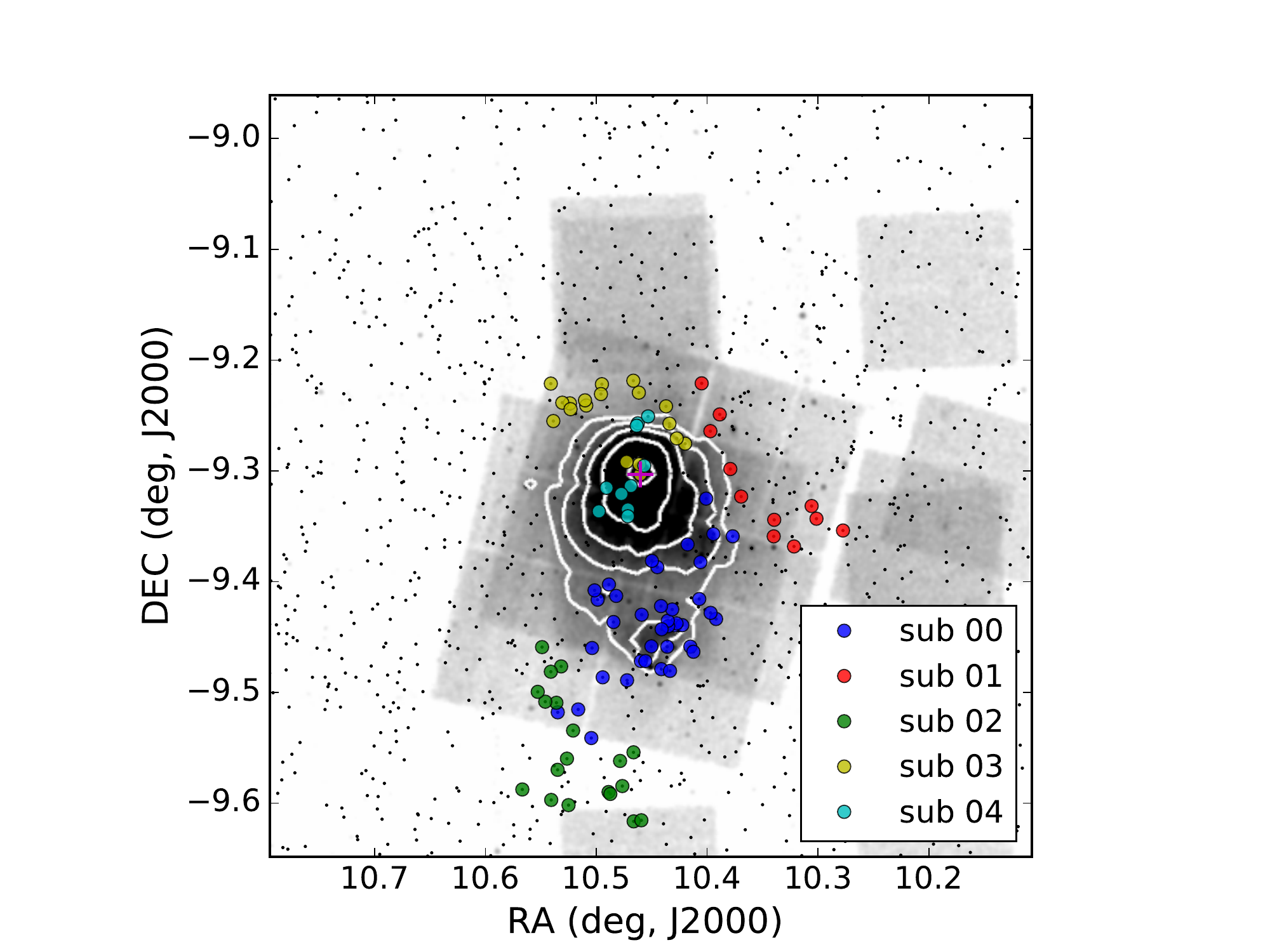}}
\caption{Substructures of the galaxy distributions overlaid on the X-ray image 
and the X-ray surface brightness contours. Colored solid circles show the galaxies belonging to the individual substructures listed in Table \ref{table:subs}. 
The position of the BCG is indicated by the purple cross. The color code is the same as Figure \ref{fig:tree}.}
\label{fig:sky}
\end{figure*}

Table \ref{table:subs} lists the 5 substructures (from sub0 to sub4) identified by 
the second threshold located at the end of the plateau at 950 km~s$^{-1}$. Table \ref{table:subs} 
also lists the mean redshift $z_o$ of the members of the substructure with its uncertainty
$\sigma_z = \sqrt{\sigma_{std}^2 + \sigma_{sp}^2}$, 
where $\sigma^2_{std}=\Sigma_i (z_i-z_o)^2/(N-1)$ is the width of the velocity
distribution of the structure members, and $\sigma_{sp}$ is the individual redshift
uncertainty.
These substructures are shown in the left panel of Figure \ref{fig:sky}
and are mainly in the cluster outskirts. Figure \ref{fig:sky}a shows that only sub0 
is associated to a clump of the X-ray emission. Sub1 to sub4 are surrounding structures.
Sub3 and sub4 coincide with the two substructures identified by \citet{2010Aguerri} 
with the \citet{1988Dressler} method based on SDSS-DR4 data alone.
We will not discuss these substructures found with the second threshold, because we do not have X-ray data 
from their corresponding area on the sky.

\begin{figure}[htbp]
\includegraphics[width=0.45\textwidth]{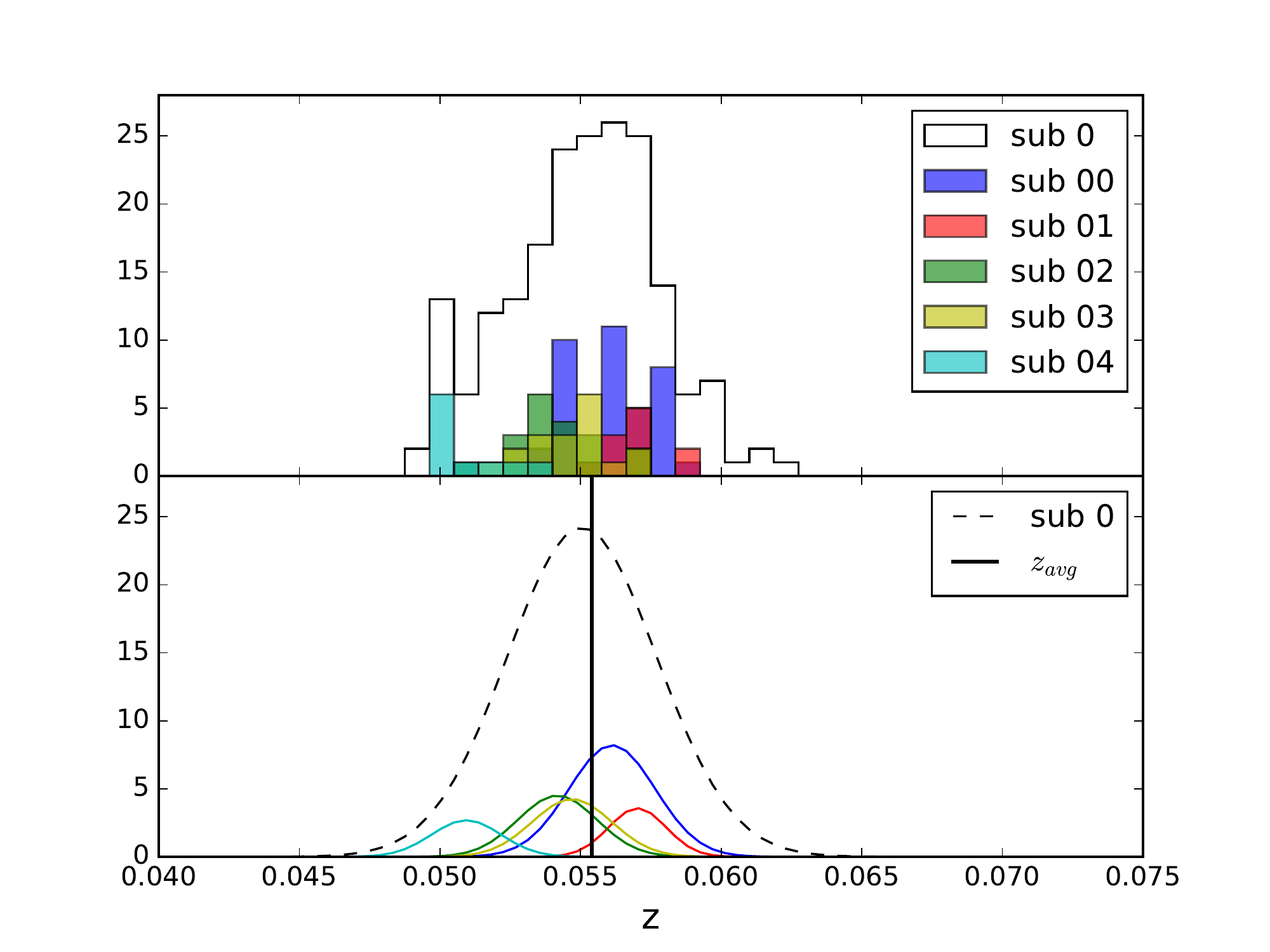}
\caption{The velocity distributions of the A85 substructures identified with the third threshold. 
The upper panel shows the velocity histograms.
The bottom panel shows the best Gaussian fits.
The black vertical line shows the position of the 
average redshift $z_{avg}=0.0554$ as a reference.}
\label{fig:vlos}
\end{figure}

\begin{table}
\centering
 \caption{Substructures of the Galaxy Distribution}
\label{table:subs}
 \begin{tabular}{|c|c|c|c|c|}
 \hline
  & $N_{gal}$ & z$_{o}$  & $\sigma_z$ & $v_{\rm disp}\;$ (km~s$^{-1})$ \\ \hline
 Cluster & 494 & 0.0554 & 0.0039 & 1054 \\
 \hline
 sub0  & 207 & 0.0546 & 0.0032 & 946 \\
 sub1  & 14  & 0.0538 & 0.0020 & 314 \\
 sub2  & 12  & 0.0586 & 0.0019 & 255 \\
 sub3  & 15  & 0.0550 & 0.0019 & 286 \\
 sub4  & 10  & 0.0582 & 0.0022 & 411 \\
 \hline
 sub00  & 38 & 0.0561 & 0.0023 & 461 \\
 sub01  & 11 & 0.0570 & 0.0020 & 307 \\
 sub02  & 19 & 0.0542 & 0.0022 & 420 \\
 sub03  & 18 & 0.0547 & 0.0022 & 423 \\
 sub04  & 10 & 0.0509 & 0.0021 & 371 \\
 \hline
 \end{tabular}
 \end{table}

The third threshold separates sub0 into individual galaxies and into the 5 substructures listed in Table \ref{table:subs} from sub00 to sub04 and
shown in the right panel of Figure \ref{fig:sky}.
The spatial distribution of these five substructures overlaps with the distribution of the X-ray emission.
The radial velocities of sub01 and sub04 show obvious discrepancies with the mean redshift (Figure \ref{fig:vlos}). 
The substructure sub00, which has the largest number of members (blue circles in the
right panel of Figure \ref{fig:sky}), has redshift $z=0.0561\pm 0.0023$, slightly
larger than the redshift of the system $z_{avg}=0.0554$, but consistent with the redshift of the X-ray sub-peak ID12,
as we will see below. In fact, sub00 is located south of the X-ray peak, but it is the counterpart of two X-ray sub-peaks. 
Sub01 (red circles) corresponds to another faint X-ray sub-peak. 
Sub02 (dark green circles) lies at the location of an X-ray filament \citep{2003Durret}.
Sub03 (yellow circles), NE (top left) of the cluster core, includes the BCG.
Sub04 (cyan circles) sits on the X-ray peak; however, its redshift $z=0.0509$ is substantially lower than $z_{avg}=0.0554$ and
the BCG redshift, but similar to the redshift of the X-ray peak ID0. 

\begin{figure}[htbp]
\includegraphics[width=0.45\textwidth]{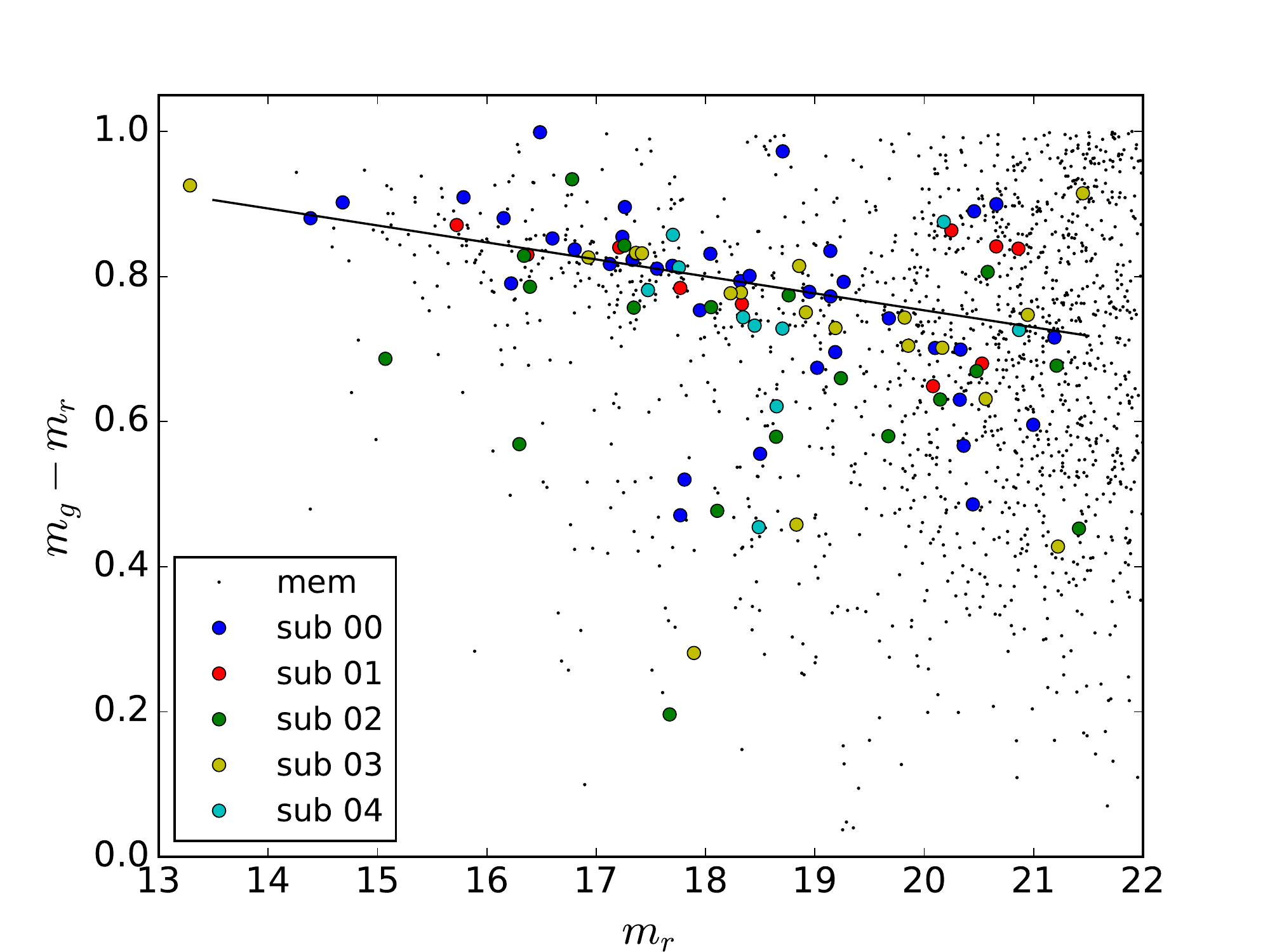}
\caption{The colour-magnitude diagram of the galaxy members of the A85 substructures identified
with the third threshold. 
The color code of the substructure is the same as in Figure \ref{fig:tree}.}
\label{fig:cmap}
\end{figure}

Using the photometric data of SDSS-DR6 \citep{2008Adelman}, 
we plot the red sequence of our substructures in Figure \ref{fig:cmap}: 
all the substructures follow the sequence of the main cluster. The fitted red sequence of the cluster is $m_g-m_r = -0.0234 m_r+1.2214$. For more details, see \citet{2016Agulli}. The outliers mainly originate 
from substructure sub02 overlapping the filament in the cluster outskirt. 

\section{Substructures in the X-ray emission}
\label{sec:xdata}

We use {\sl Chandra} archived data to estimate the redshift of the cluster 
and its substructures with the X-ray
spectrum fitting procedure. 
The ObsID 904 is 16 years old and in a different observation mode.
The ObsIDs from 4881 to 4888, that contain the A85 field of view, are shallow and offset and do not
contain any recognizable structures. 
To avoid possible calibration errors, we adopt the four most recent ObsIDs, 
taken in 2013, listed in Table \ref{xlist}. 

The selected observations were carried out between August 9$^{\rm th}$ and 17$^{\rm th}$ 2013 in 
VFAINT mode using the Advanced CCD Imaging Spectrometer (ACIS-I). 
The data reduction is performed using the latest release of
the {\sl ciao} software (version 4.8) with CALDB 4.7.0. 
The charge transfer inefficiency (CTI) correction, time-dependent gain adjustment, grade correction,
and pixel randomization are applied.  We are able to filter efficiently the background events 
thanks to the VFAINT mode, thus reducing the background by $\sim 25$\% in the 
hard (2.0-10 keV) band.  Eventually, we search for high background spikes, 
and remove them with a 3$\sigma$ clipping.  The final exposure times
of each ObsID are lower than the nominal exposure time only by a few percent.
The {\sl level 2} event files obtained in this way are reprojected to match the coordinates of ObsID 15173, 
and merged into a single event file. The total exposure time of the merged data is $\sim 156.7$ ks.

\begin{table}
\centering
 \caption{List of Chandra observations}
\label{xlist}
 \begin{tabular}{|c|c|c|c|c|}
 \hline 
 ObsID & Exp (ks) & chips & Mode & Date \\  \hline 
 15173 & 42.52 & ACIS-I & VFAINT & 2013-08-14 \\
 15174 & 39.55 & ACIS-I & VFAINT & 2013-08-09 \\
 16263 & 38.15 & ACIS-I & VFAINT & 2013-08-10 \\
 16264 & 36.6 & ACIS-I & VFAINT & 2013-08-17 \\ \hline 
\end{tabular}

\end{table}
%
%
%

\begin{figure}[htbp]
\includegraphics[width=0.45\textwidth]{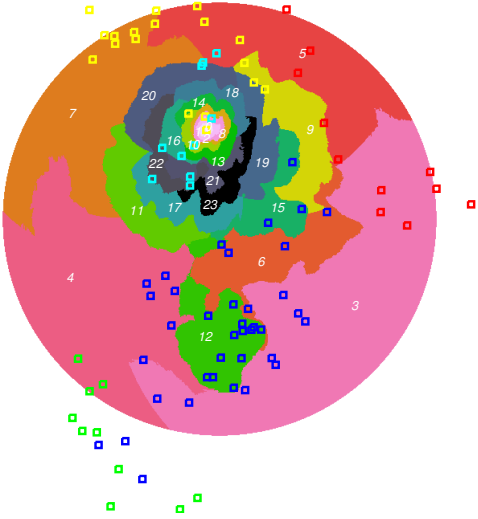}
\caption{The region division of the X-ray surface brightness with 
the substructures of the galaxy distribution overlaid. 
The colors of the galaxies (squares) are the same as in the right panel of Figure \ref{fig:sky}.}
\label{fig:xreg}
\end{figure}

Similarly to the analysis performed by \citet{2015Liu} on the Bullet cluster and other clusters \citep{2016Liu}, 
we apply the contour-binning technique of \citet{Sanders2006} to select 
regions according to the surface brightness distribution of an
extended source. Our goal is to obtain spectra with comparable
quality for the measurement of the X-ray redshift. To achieve this goal, we
require comparable numbers of net counts in the
0.5-8 keV band in all the regions. We select a circular region of 530 arcsec
including all the visible structures within the chips. This circular region is divided into
24 regions with the condition that the full-band signal-to-noise ratio (S/N) is larger 
than 200 in each region. 
The location and the ID number of these regions are shown in Figure \ref{fig:xreg}.
There are two sub-peaks around the cluster: the peak in region 12 (sub-peak A) at the bottom of the image is bright;
the peak in region 15 (sub-peak B) at the right of the image is relatively faint.

Because the cluster covers almost the entire X-ray field,
we use a nearby region out of our target region to remove the background.
All spectra are fitted with {\sl Xspec} v12.9.0i \citep{1996Arnaud}
in the full band ($0.5-8.0$ keV).
To model the X-ray emission, we use double {\tt mekal} plasma emission models 
\citep{1985Mewe, 1986Mewe,1992Kaastra,1995Liedahl} which include thermal bremsstrahlung and 
line emission, with abundances measured relatively to the solar values in 
\citet{2005Asplund}, where Fe/H $ = 3.6 \times 10^{-5}$.
The double-temperature thermal spectrum is helpful to reduce the possible bias in the measurement of
the iron line centroid due to the presence of unnoticed thermal structure along the line of sight.
Galactic absorption is described by the model {\tt tbabs} \citep{2000Wilms}. 
The ICM temperature,  the heavy elements abundance, the X-ray redshift 
and the normalization parameter are all set unconstrained at the same time.
The redshifts of the two-temperature components are always the same.
Considering that there is a large parameter space to explore, we adopt the Monte Carlo
Markov Chain (MCMC) method to fit the spectrum. The chain is generated by
the Goodman-Weare algorithm \citep{goodman2010}, with 10 walkers, 10000 burn steps
and the total length of 1000000 steps.
After the fitting, chains are top-hat filtered according to the following ranges: 
temperature from 0 to 20 keV,
metallicity from 0 to 2, and redshift from 0 to 0.1.
The best fitting parameters and their errors are estimated from these filtered chains.

\begin{figure}[htbp]
\includegraphics[width=0.45\textwidth]{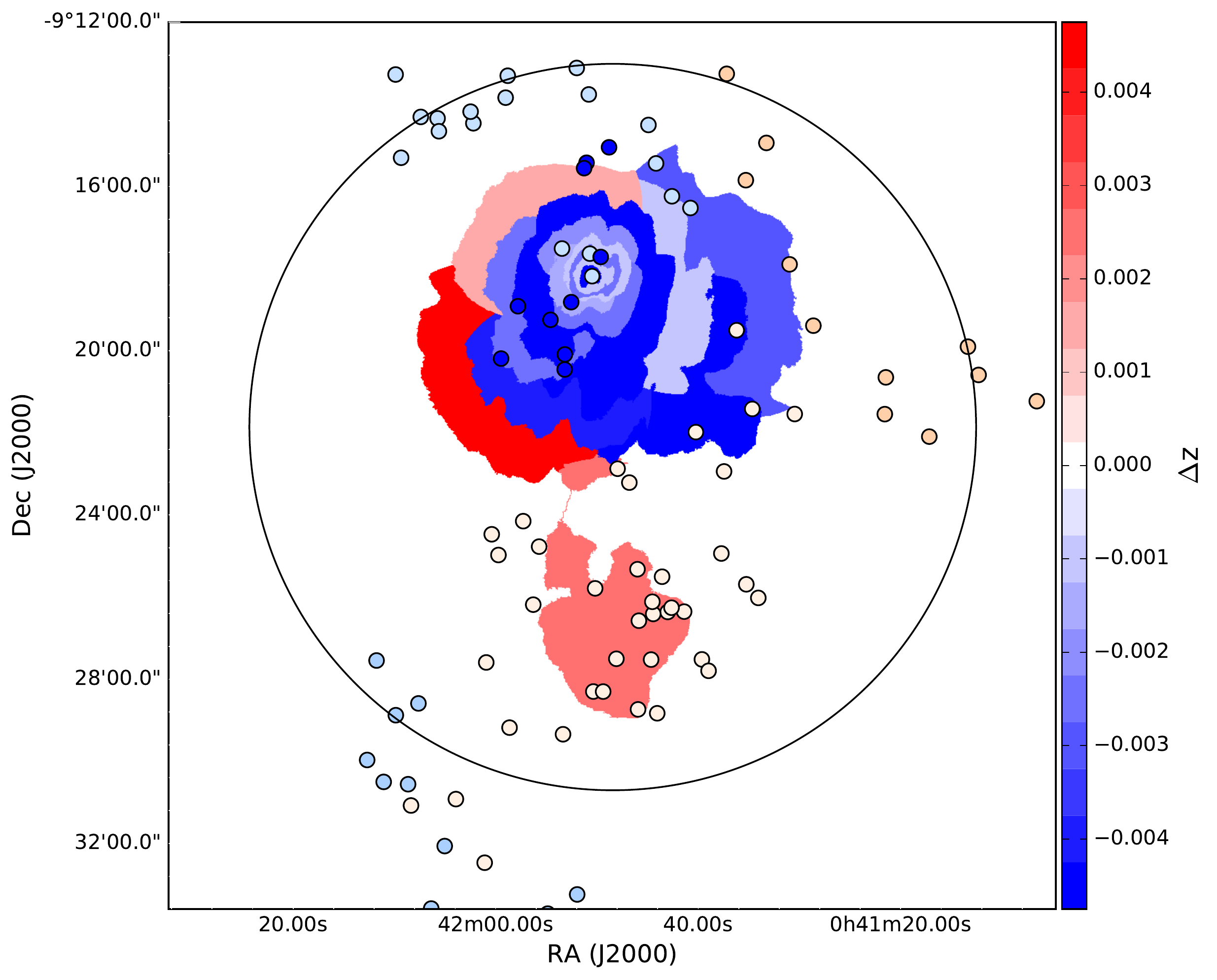}
\caption{Redshift deviation map of the ICM of A85 with respect to its average 
redshift ($z_{avg}=0.0554$). The bin size of the colored bar is $\delta z = 0.001$.
The galaxies (circles) are colored with the same redshift scale.
}
\label{fig:zmap}
\end{figure}

As we mentioned above, the stacked X-ray image is divided into 24 regions.
The fitting results of all the regions are listed in Table \ref{fit}.
Figure \ref{fig:zmap} shows the redshift deviation ($z_X-z_{avg}$) map with respect to the average 
redshift.
Because the X-ray redshift error strongly depends on metallicity,
regions ID 3, 4, 5 and 7 in the cluster outskirts of Figure \ref{fig:xreg} 
do not have effective redshift values due to their low metallicity ($Z/Z_\odot< 0.15$). 
Region ID 0 in the cluster center 
has the most precise redshift because of its high metallicity.

With these measurements, we measure a redshift deviation between the 
core and the X-ray substructures, which is consistent with the results of \citet{2005Durret} 
obtained with {\sl XMM} data. In addition, for the first time, we can now compare 
the redshifts of these regions with those of the optical substructures, 
to investigate the redshift structure of the cluster.

\begin{table}
\centering
 \caption{X-ray Region Fitting Results}
\label{fit}
\begin{tabular}{|c|c|c|c|c|c|}
 \hline
ID & $T_1$ & $Z_1$ & $T_2$ & $Z_2$ & z \\
\hline
0 & $3.47_{-0.18}^{+5.80}$ & $1.16_{-1.06}^{+0.12}$ & $3.47_{-0.48}^{+0.12}$ & $1.10_{-0.08}^{+0.47}$ & $0.0501_{-0.0013}^{+0.0014}$ \\
1 & $5.01_{-0.74}^{+0.59}$ & $0.46_{-0.21}^{+0.34}$ & $4.29_{-0.37}^{+0.26}$ & $1.16_{-0.34}^{+0.53}$ & $0.0548_{-0.0017}^{+0.0014}$ \\
2 & $4.98_{-0.42}^{+0.96}$ & $0.61_{-0.42}^{+0.62}$ & $5.14_{-0.51}^{+1.02}$ & $0.57_{-0.41}^{+0.59}$ & $0.0533_{-0.0015}^{+0.0016}$ \\
3 & $4.48_{-0.31}^{+0.70}$ & $0.04_{-0.03}^{+0.05}$ & $5.52_{-0.93}^{+0.41}$ & $0.12_{-0.05}^{+0.05}$ & - \\
4 & $6.55_{-0.53}^{+0.74}$ & $0.00_{-0.0}^{+0.0}$ & $6.26_{-0.74}^{+0.97}$ & $0.13_{-0.09}^{+0.095}$ & - \\
5 & $3.76_{-0.21}^{+0.78}$ & $0.03_{-0.03}^{+0.17}$ & $6.70_{-2.01}^{+0.23}$ & $0.02_{-0.01}^{+0.02}$ & - \\
6 & $5.85_{-0.58}^{+0.50}$ & $0.19_{-0.06}^{+0.02}$ & $6.05_{-0.58}^{+2.80}$ & $0.17_{-0.06}^{+0.40}$ & $0.0556_{-0.0039}^{+0.005}$ \\
7 & $6.89_{-0.47}^{+0.73}$ & $0.02_{-0.01}^{+0.02}$ & $6.75_{-0.89}^{+2.58}$ & $0.01_{-0.01}^{+0.01}$ & - \\
8 & $5.24_{-0.54}^{+0.83}$ & $0.64_{-0.34}^{+0.72}$ & $5.42_{-0.63}^{+0.83}$ & $0.54_{-0.32}^{+0.51}$ & $0.0546_{-0.0014}^{+0.0011}$ \\
9 & $7.08_{-1.15}^{+1.61}$ & $0.16_{-0.09}^{+0.58}$ & $7.68_{-0.89}^{+0.32}$ & $0.19_{-0.12}^{+0.37}$ & $0.0527_{-0.0043}^{+0.0057}$ \\
10 & $6.50_{-1.25}^{+3.12}$ & $0.39_{-0.28}^{+0.44}$ & $5.52_{-0.61}^{+0.88}$ & $0.51_{-0.17}^{+0.32}$ & $0.0542_{-0.0018}^{+0.0023}$ \\
11 & $7.12_{-1.42}^{+2.19}$ & $0.24_{-0.18}^{+0.04}$ & $6.08_{-1.21}^{+3.13}$ & $0.28_{-0.13}^{+0.23}$ & $0.0611_{-0.0037}^{+0.003}$ \\
12 & $5.64_{-1.87}^{+2.02}$ & $0.27_{-0.21}^{+1.26}$ & $5.28_{-1.45}^{+2.01}$ & $0.35_{-0.27}^{+0.46}$ & $0.0583_{-0.0047}^{+0.0039}$ \\
13 & $6.15_{-0.69}^{+2.07}$ & $0.59_{-0.16}^{+0.24}$ & $6.51_{-0.49}^{+2.20}$ & $0.37_{-0.30}^{+0.08}$ & $0.0532_{-0.0014}^{+0.0028}$ \\
14 & $5.95_{-0.72}^{+0.89}$ & $0.41_{-0.29}^{+0.48}$ & $5.43_{-0.73}^{+1.08}$ & $0.73_{-0.62}^{+0.38}$ & $0.0538_{-0.0016}^{+0.0018}$ \\
15 & $7.16_{-2.89}^{+2.81}$ & $0.22_{-0.12}^{+0.27}$ & $6.68_{-1.30}^{+2.38}$ & $0.40_{-0.13}^{+0.13}$ & $0.0493_{-0.0058}^{+0.0036}$ \\
16 & $6.63_{-2.03}^{+3.19}$ & $0.88_{-0.33}^{+0.41}$ & $5.62_{-1.06}^{+0.42}$ & $0.41_{-0.10}^{+0.07}$ & $0.0499_{-0.0015}^{+0.0022}$ \\
17 & $5.78_{-1.86}^{+0.76}$ & $0.40_{-0.21}^{+0.58}$ & $6.86_{-0.69}^{+1.60}$ & $0.35_{-0.10}^{+0.39}$ & $0.0519_{-0.0012}^{+0.0013}$ \\
18 & $5.93_{-0.55}^{+1.05}$ & $0.32_{-0.25}^{+0.20}$ & $5.66_{-0.80}^{+4.12}$ & $0.91_{-0.53}^{+0.18}$ & $0.0500_{-0.0016}^{+0.0019}$ \\
19 & $7.39_{-1.23}^{+2.24}$ & $0.20_{-0.15}^{+0.16}$ & $7.08_{-1.34}^{+2.24}$ & $0.36_{-0.16}^{+0.39}$ & $0.0549_{-0.0034}^{+0.0035}$ \\
20 & $7.21_{-1.80}^{+1.68}$ & $0.25_{-0.17}^{+0.72}$ & $6.75_{-1.42}^{+0.89}$ & $0.22_{-0.16}^{+0.48}$ & $0.0571_{-0.0042}^{+0.0044}$ \\
21 & $6.75_{-0.73}^{+1.00}$ & $0.49_{-0.12}^{+0.13}$ & $6.07_{-1.25}^{+3.18}$ & $0.37_{-0.17}^{+0.33}$ & $0.0496_{-0.0023}^{+0.0023}$ \\
22 & $8.71_{-2.40}^{+3.55}$ & $1.93_{-0.13}^{+0.28}$ & $5.92_{-0.39}^{+0.21}$ & $0.31_{-0.08}^{+0.08}$ & $0.0532_{-0.0019}^{+0.0023}$ \\
23 & $6.97_{-1.27}^{+2.19}$ & $0.13_{-0.10}^{+1.47}$ & $6.98_{-0.85}^{+0.81}$ & $0.41_{-0.37}^{+0.35}$ & $0.0485_{-0.0028}^{+0.0038}$ \\
\hline
\end{tabular}
\end{table}

\section{Combining the optical and X-ray spectroscopy}
\label{sec:comb}
\begin{figure}[htbp]
\includegraphics[width=0.45\textwidth]{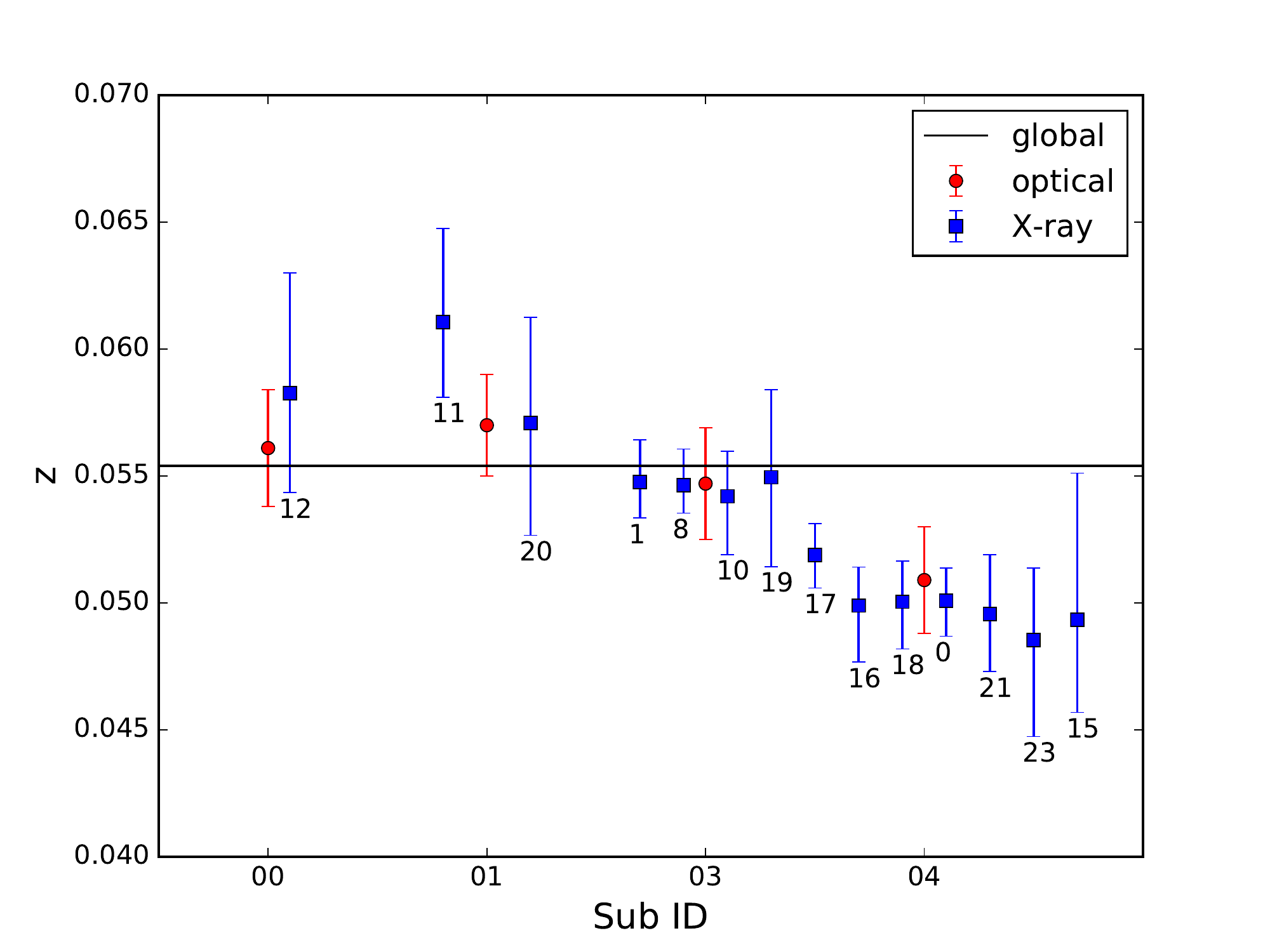}
\caption{The redshifts of the optical substructures of A85 (red dots) and the redshifts 
of the X-ray regions (blue squares). 
The abscissas of the blue squares are chosen accordingly to their supposedly 
correlated optical substructures. 
The black solid line is the mean redshift $z_{avg}=0.0554$.}
\label{fig:comb}
\end{figure}

\begin{figure}[htbp]
\includegraphics[width=0.45\textwidth]{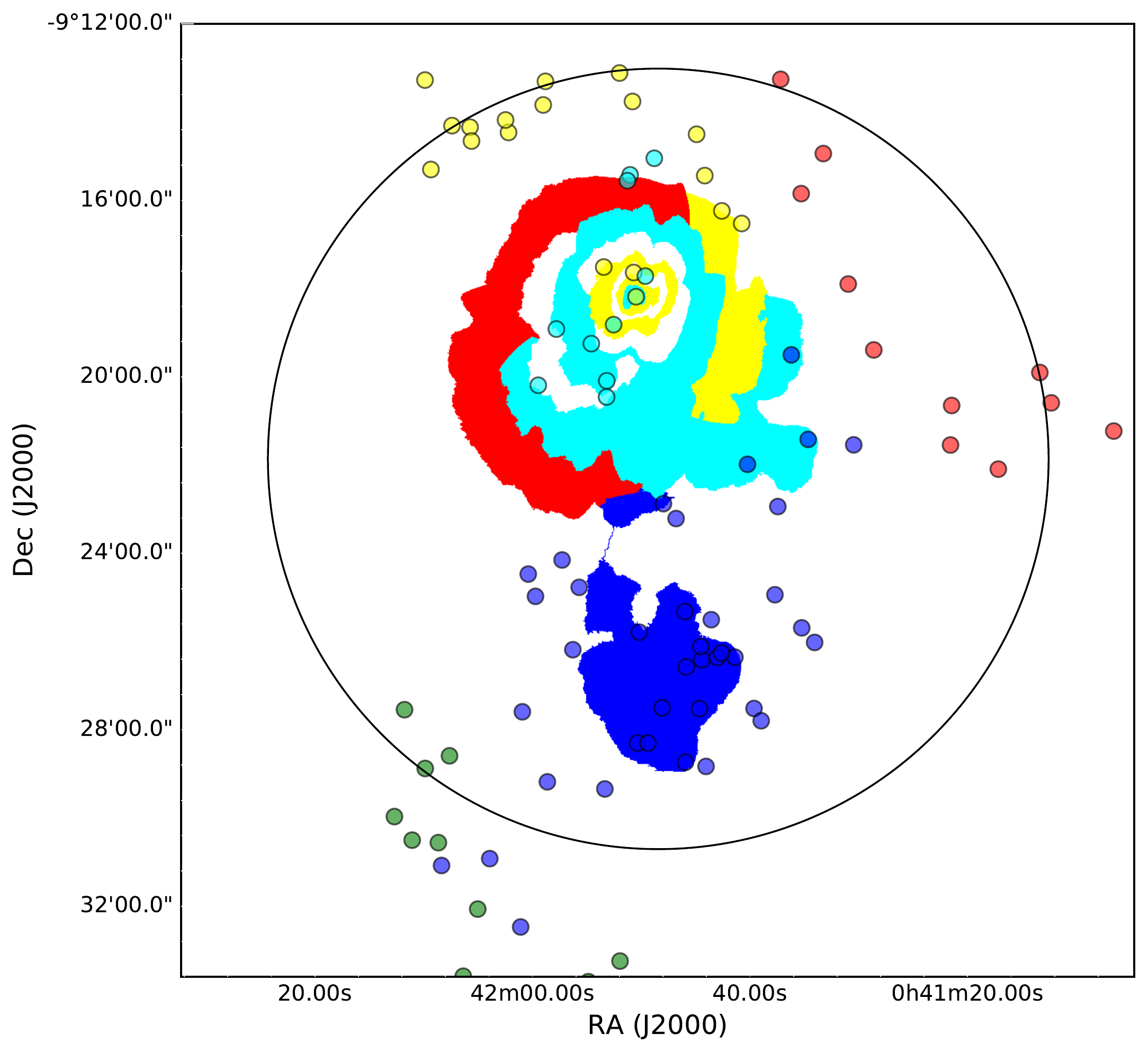}
\caption{The spatial distribution of the associated optical substructures and 
the X-ray regions. The related regions share the same color. }
\label{fig:fitsub}
\end{figure}

In this section, we combine the X-ray results with the properties of 
the optical substructures sub00 to sub04, located
in the central region of the cluster, identified with the third threshold
on the binary tree of the caustic technique. We do not discuss sub02, because it is out of the Chandra field. 
The associations between galaxy substructures and ICM that we describe below are summarized in Figures \ref{fig:comb} and \ref{fig:fitsub}.

The mean position of sub04 (cyan circles in Figure \ref{fig:fitsub}) overlaps with the central peak of the X-ray emission (Figure \ref{fig:xreg}), 
with redshift $z=0.0509\pm 0.0021$; 
the X-ray central 
regions ID 2, 13, 14, and 22 (yellow areas in Figure \ref{fig:fitsub}) have redshifts in the range $0.0532-0.0538$. These 
redshifts appear substantially different from the sub04 redshift, 
suggesting that sub04 is now passing through the cluster core along the line of sight 
towards the observer at a velocity  $v\sim c(0.0535-0.0509)=780$~km~s$^{-1}$, dragging some of the
core X-ray gas, as indicated by the redshift of ID 0, $z=0.0501^{+0.0014}_{-0.0013}$, consistent with the redshift of sub04. 
The substructure sub04 is falling into A85 with its own gas that can be identified with the X-ray regions 
ID 15, 16, 17, 18, 21, and 23 (cyan areas in Figure \ref{fig:fitsub}), whose redshifts are in the range $0.0485-0.0519$, whereas some of its gas (ID 9, $z=0.0527^{+0.0057}_{-0.0043}$) might be 
lagging behind because of ram pressure.  
In fact, galaxies are collisionless objects in cluster dynamics, 
and, during the cluster merging, they can easily move with the 
substructure of dark matter and separate from the baryonic gas 
\citep{2004Markevitch,2008Bradac,2011Merten,2012Dawson,2013Dahle,2014Gastaldello}.

Another group, sub03 (yellow circles in Figure \ref{fig:fitsub}), with redshift $z=0.0547\pm 0.0022$, is going through the cluster
from SW (bottom right of the plot of Figure \ref{fig:fitsub}) to NE (top left of the plot) 
almost perpendicularly to the line of sight. Unlike sub04, sub03 is not going exactly through the 
cluster core. Similarly to sub04 however, sub03 is moving with its own gas, associated to the X-ray regions ID 1, 8, 10, and 19 (yellow areas in Figure \ref{fig:fitsub}), with 
redshifts in the range $0.0542-0.0549$, and it might be leaving behind some gas associated to ID 6 ($z=0.0556^{+0.0050}_{-0.0039}$). 
The BCG, whose redshift is $z=0.0554\pm 0.0002$, belongs to sub03 and is located at the end of a plume of the 
X-ray temperature map recently unveiled \citep{2015Ichinohe}, suggesting that it is indeed moving through the cluster, exactly
like sub03.

Sub01 (red circles in Figure \ref{fig:fitsub}), 
at redshift $z=0.0570\pm 0.0020$, which is looser and poorer than the other substructures, 
also appears to be going through the cluster
from E (left of the plot of Figure \ref{fig:fitsub}) 
to W (right of the plot) and is leaving behind some of its own gas associated
to the region ID 20 (part of the red area in Figure \ref{fig:fitsub}), with $z=0.0571^{+0.0044}_{-0.0042}$, and probably also the region ID 11 (part of the red area in Figure \ref{fig:fitsub}),
with $z=0.0611^{+0.0030}_{-0.0037}$, if sub01 has a non-negligible velocity component towards
the observer.  


Sub00 (blue circles in Figure \ref{fig:fitsub}) is the largest substructure in the cluster, covers a large area, and overlaps with two X-ray sub-peaks. 
Its redshift, $z=0.0561\pm 0.0023$, 
is consistent with the redshift of the X-ray sub-peak associated
to region ID 12 (blue area in Figure \ref{fig:fitsub}),  $z=0.0583^{+0.0039}_{-0.0047}$. 

\section{Discussion and Conclusion}
\label{sec:resu}

For the first time, we combine two techniques, 
the caustic technique based on optical spectroscopic redshifts and the X-ray redshift fitting procedure, 
to explore the complex redshift structures of the central region of a galaxy cluster, A85,
within $r_{500}\sim 0.6h^{-1}$~Mpc, where the two probes provide overlapping data.

The substructures in the galaxy distribution and the ICM dishomogeneities observed in the X-ray 
band are correlated 
with each other, although they do not always share the same redshift and position on the sky.
Galaxies and ICM have different evolving time scale, and are usually in different phases during a merging event.

We identify five substructures in A85 within $\sim 600 h^{-1}$~kpc from the cluster X-ray peak: 
two substructures, sub01 and sub03, appear to have been recently accreted, and 
a substructure, sub00, is being currently accreted; the last substructure, sub04, with
an optical redshift $z=0.0509$, has just gone through   
the cluster core almost exactly along the line of sight and is now
moving out of the cluster towards the observer. Its existence is the main reason why 
previous investigations, both in the optical and X-ray, measured a cluster  
redshift smaller than the average redshift $z_{avg}=0.0554$. 
We conclude that A85 is not a relaxed system but has experienced recent merging events, 
in agreement with other investigations (e.g., \citealt{2015Ichinohe}).
It will be interesting to investigate whether
an accurate $N$-body/hydrodynamical simulation of the cluster 
mass accretion can reproduce the optical and X-ray kinematic properties of 
A85 that we find here. 

Our analysis rests on a sample of optical redshifts mostly ($\sim 85\%$) derived from low-resolution spectra
with $\sim 500$~km~s$^{-1}$ uncertainty. This uncertainty is comparable to the cluster velocity
dispersion $692^{+55}_{-45}$~km~s$^{-1}$ \citep{Rines2006} and might affect the solidity of our substructure
identification. 

\citet{Yu2015} show that analyses of mock catalogs with perfectly known redshifts, based on
the caustic technique with the automatic plateau and threshold identifications, 
return substructure catalogs that are $\sim 60$\% complete and contain $\sim 50$\% spurious substructures; 
\citet{Yu2015} also suggest that these results can improve when the binary tree threshold
is tuned by hand, as we do here. 

We tested our results as follows: we create synthetic
redshift samples by replacing each individual redshift with a random variate extracted
from a Gaussian probability density distribution with its mean set by the measured redshift
and its dispersion set by the redshift uncertainty. Although the substructures identified 
in these synthetic samples might differ from sample to sample, we can still
infer the same general picture: specifically we always find substructures associated
with the X-ray filament and sub-peaks. The existence of the substructures called here sub03 and sub04 requires
to be confirmed by more precise redshifts derived from high-resolution optical spectra.
Finally, the uncertainties of the X-ray redshifts, which are the smallest uncertainties
that can be obtained with current instrumentation, are $\sim 400$~km~s$^{-1}$ or larger; therefore,
our comparison is based on optical and X-ray data with comparable uncertainty, but clearly more robust
results can only be reached by improving both optical observations and X-ray technology.    

There are some uncertainties and limitations both in the substructure identification with
the caustic method and in the X-ray spectrum fitting procedure. 
Even with perfectly know individual galaxy redshifts, projection effects unavoidably 
weaken the solidity of the identification of the substructures with the caustic method 
and the estimate of their properties and their uncertainties, including their mean redshifts.
On the other hand, the energy resolution of the X-ray detectors are limited and it appears  
hard to improve the precision of the redshift measurement with current devices.

Before the advent of a bolometer with high angular resolution, similar to 
what is planned for the upcoming X-ray telescope ATHENA\citep{Athena}, that will certainly provide a 
dramatic improvement in the field, an intermediate mission with instrumental features
similar to the unfortunate X-ray observatory Hitomi \citep[Astro-H,][]{AstroH}
will be invaluable for the detailed investigation of the dynamics and
thermodynamics of the ICM \citep{2016Hitomi}.

When extended to a large cluster sample, the combined analysis 
of dense optical redshift surveys with those improved X-ray spectroscopy 
from future X-ray telescopes, 
will certainly provide an essential contribution to our
understanding of the growth history of galaxy clusters.

\acknowledgments
We sincerely thank Margaret Geller, Ana Laura Serra, and Jubee Sohn for numerous fruitful discussions.
 We acknowledge support from the grant Progetti di
Ateneo/CSP$\_$TO$\_$Call2$\_$2012$\_$0011 ``Marco Polo'' of the University of Torino, 
the INFN grant InDark, the grant PRIN 2012 ``Fisica Astroparticellare Teorica'' of the Italian
Ministry of University and Research, the National Natural Science Foundation of China 
under Grants Nos. 11373014, 11073005, and 11403002,
the Fundamental Research Funds for the Central Universities and Scientific Research Foundation 
of Beijing Normal University.
This work has been partially funded by the MINECO (grant AYA2013-43188-P).
This research has made use of the Sixth Data Release of SDSS, and of the 
NASA/IPAC Extragalactic Database which is operated by the Jet Propulsion 
Laboratory, California Institute of Technology, under contract with the 
National Aeronautics and Space Administration.
The WHT and its service programme are operated on the island of La Palma 
by the Isaac Newton Group in the Spanish Observatorio del Roque de los 
Muchachos of the Instituto de Astrof\'isica de Canarias.
IA and JALA acknowledge support from the Ministerio de Economia y Competitividad (MINECO) under the grant AYA2013-43188-P.
P.T. is supported by the Recruitment Program of High-end Foreign
Experts and he gratefully acknowledges hospitality of Beijing Normal
University.

\bibliography{a85}
\end{document}